# On the retrieval of attenuation from the azimuthally averaged coherency of a diffuse field

Richard L Weaver
Department of Physics
University of Illinois
Urbana, Illinois 61801

Abstract

It has been suggested that seismic attenuation $\alpha$ can be inferred from comparisons of empirical coherencies (the cross spectra of pre-whitened ambient seismic noise records) with attenuated Bessel functions $J_o(\omega r/c) \exp(-\alpha\, r)$. Analysis shows here, however, that coherency depends strongly on the directionality of ambient noise intensity. Even if coherency is azimuthally averaged, the suggested attenuation dependence $\exp(-\alpha r)$ does not apply. Indeed in highly directional noise fields, coherency is independent of attenuation. It is also argued that spatial and azimuthal averages of coherency can incur phase cancellations related to variations in wavespeed that may mimic factors like $\exp(-\alpha r)$. Inference of attenuation from comparison of empirical coherencies to $J_o(\omega r/c) \exp(-\alpha\, r)$ is problematic.

Introduction

Retrieval of seismic velocities from the correlations of seismic noise is now well developed. The impetus for this development may be traced to the observation (Weaver and Lobkis 2001, Lobkis and Weaver 2001, Weaver and Lobkis 2004) that fully equipartitioned diffuse waves permit, on cross-correlating the noise, retrieval of the Green function, where the Green function is the signal one would have at one receiver if the other were replaced with an impulsive force. While seismic noise fields are rarely fully equipartitioned, it later transpired that a similar relation obtains under more general conditions. These conditions include a smooth directional distribution of ambient noise intensity, and station separation large compared to a wavelength [Snieder 2004, Froment *et al.* (2010), Godin (2009), Weaver *et al.* (2009.)] The modified relation has proved sufficient for extracting wave velocities, but not attenuation.

The possibility of interpreting correlation amplitudes as well as arrival times, with a view towards retrieving seismic attenuation, site amplification factors, and scattering strengths is now attracting attention [ Matzel 2007, Prieto and Beroza 2008, Prieto et al 2009, Harmon et al 2010, Prieto et al 2011, Lawrence and Prieto 2011, Tsai 2011, Lin et al 2011, Weaver 2011a, 2011b]. It has become abundantly clear that noise directionality, or as it is sometimes called, "ponderosity," plays an important role. Weaver (2011b) suggests that correlation amplitudes can be interpreted and attenuation retrieved by modeling ambient noise directionality, especially how it varies in space and azimuth. The suggestion was supported by numerical experiments

A very different approach has been pursued by others (Prieto *et al.* 2009, 2011, Lawrence and Prieto 2011) who use azimuthal and regional averages (at fixed detector separation *r*) of correlations in the frequency domain. Before that averaging, they normalize the correlations between each pair of stations by the rms at the two stations. They note the cross-spectrum normalized in this manner is termed the "coherency." Amongst its virtues, one notes that coherency is independent of seismic site amplification factors. It is claimed that azimuthally averaged coherency can be compared to a Bessel function $J_o(\omega r/c)$ times an exponential diminishment $\exp(-\alpha\ r)$, and from this comparison (average) wavespeed $c$ and (average) attenuation $\alpha$ may be extracted. The particular virtue of the azimuthal averaging is that it may render the ambient field effectively isotropic and obviate concerns over noise directionality. It is also conjectured that such averaging mitigates the effects of focusing and defocusing. That this correlation ought look like a Bessel function $J_o$ is well established (Aki, 1957.) That deviations from a Bessel function ought be ascribable to a factor $\exp(-\alpha r)$ is less clear. Nevertheless, impressive maps of seismic anelasticity have been developed.

Recently Tsai(2011) has examined coherency theoretically with and without azimuthal averaging and concluded that, while attenuation plays an important role, so does the distribution of sources. His expressions show that an azimuthal average of coherency does not fully remove the effects of noise directionality. He concludes that identification of coherency with $J_o(\omega r/c) \exp(-\alpha\ r)$ is justified only under special circumstances.

Here we examine the case further. It is argued that identification of azimuthally averaged coherency with $J_o(\omega r/c) \exp(-\alpha\ r)$ for the retrieval of attenuation $\alpha$ is not justified. The radial decay is not exponential; as Tsai showed, it depends on the angular dependence of intensity. Inasmuch as that necessarily varies with position in an attenuating medium, it also depends on position. It is further argued that the radial decay of coherency is affected by wavespeed anisotropy and/or spatial variations in wavespeed.

Coherency

Prieto et al (2009, 2011) and Lawrence and Prieto(2011) have suggested that the *coherency* between signals at two positions $\vec{x}$ and $\vec{o}$, especially if it is averaged over azimuth and over position, may be analyzed for the purposes of retrieving attenuation. Coherency is defined as the normalized cross spectrum of two frequency domain signals

$$C^{(1)} \equiv \frac{\psi_o^*(\omega)\psi_{\vec{x}}(\omega)}{|\psi_o(\omega)||\psi_{\vec{x}}(\omega)|} \tag{1}$$

where $\psi_{\vec{y}}(\omega)$ is the (Fourier transform of) the scalar signal detected at the station at position $\vec{y}$. An asterisk * indicates complex conjugation. A separate ensemble, or time, average of the numerator and denominator is understood. This construction removes the influence of the (unknown) site amplification factors, thereby eliminating them as sources of uncertainty, An inverse Fourier transform serves to define the coherency in the time domain (Prieto et al 2011).

A slightly different quantity is

$$C^{(2)} \equiv \frac{\psi_o^*(\omega)\psi_{\vec{x}}(\omega)}{|\psi_o(\omega)|^2} \tag{2}$$

in which the cross spectrum is normalized by the mean square field at the reference station. Matzel (2007) introduced that normalization for time domain correlations. $C^{(2)}$ may also be interpreted as the signal at $\vec{x}$ deconvolved on the signal at o. Prieto et al (2011) term this a transfer function. $C^{(2)}$ is *not* independent of site factors. Both normalizations are such that C at zero distance, $o = x$, is unity. $C^{(2)}$ is analyzed in the next section.

The two C are identical in a fully diffuse field for which noise is homogeneous and isotropic. (In a lossy medium, this requires that sources are distributed with strength in proportion to attenuation[Weaver 2008], a condition which is not realized in seismology.) Under those conditions the usual theorems apply and each C is merely the identically normalized imaginary part of the frequency-domain Green function. Both speed and amplitude could then be retrieved from the identification, in 2-d, C= Re($H_o^{(1)}(\omega r/c+i\alpha r)$)/Re($H_o^{(1)}(0)$) ~ $J_o(\omega r/c)$ exp($-\alpha r$) where $H_o^{(1)}$ is the Hankel function. Under these conditions both quantities C oscillate and diminish as suggested by Prieto *et al* (2009,2011) and Lawrence and Prieto (2011).

We ask what form averages of these quantities C take, and what role attenuation plays in them, for various other choices of ambient diffuse field. We first consider $C^{(1)}$, the coherency, and examine what form it takes with and without azimuthal averaging.

The simplest case is that in which the insonifying noise field is a single plane wave in the +x direction. The field at a point $\vec{x} = \{r, \theta\}$ is

$$\psi_{\vec{x}=\{r,\theta\}} = A\exp(i\omega r\cos\theta/c - \alpha r\cos\theta) \tag{3}$$

Without loss of generality the position *o* may be taken to be the origin. (Shifting it merely affects the mean square value of the random complex amplitude A.) Straightforward algebra leads to

$$C^{(1)} = \exp(i\omega r\cos\theta/c) \tag{4}$$

Remarkably, and perhaps counter-intuitively, this coherency has no dependence on attenuation. Inasmuch as the amplitude A has dropped out, this result is independent of the choice of reference position *o,* and so a spatial average at fixed *r* leaves the expression unchanged.

On azimuthally averaging over all receiver pair orientations θ, $C^{(1)}$ becomes

$$<C^{(1)}>_\theta = \frac{1}{2\pi}\int C^{(1)} d\theta = J_o(\omega r/c) \tag{5}$$

a result equivalent to that of Aki(1957).   This is independent of $\alpha$.  At least for this special case in which the ambient field is a single plane wave, one cannot retrieve attenuation from $C^{(1)}$.

These conclusions differ slightly in the case of a more arbitrary angular distribution of incident noise, i.e. for an insonifying noise field that is not a simple unidirectional plane wave.  In this case the field at $\vec{x} = \{r,\theta\}$ is written as a superposition of plane waves

$$\psi_{\vec{x}=\{r,\theta\}} = \oint d\phi \ s(\phi) \exp(i\omega r \cos(\theta - \phi)/c - \alpha r \cos(\theta - \phi)) \qquad (6)$$

where $s(\phi)$ is the complex amplitude of the wave in the $\phi$ direction evaluated at the origin.  As is conventional [e.g. Weaver et al 2009], the diffuse field is taken to have statistics such that

$$<s(\phi)s^*(\phi')> = B(\phi)\delta(\phi - \phi'), \qquad <s(\phi)s(\phi')> = 0 \qquad (7)$$

corresponding to uncorrelated plane waves with intensity at the origin equal to $B(\phi)$.  A shift in the origin must be accommodated by a corresponding change in the function $B(\phi)$.  For example, a field whose intensity at the origin is isotropic, will be weighted towards left going waves at a position to the right of the origin closer to the source of the left going waves and further from the source of the right going waves.

$C^{(1)}$ is evaluated as the ratio of expectations [ Prieto et al 2011]

$$C^{(1)} \equiv \frac{\psi_o^* \psi_{\vec{x}}}{|\psi_o||\psi_x|} = \frac{\oint B(\phi)\exp(i\omega r \cos(\theta - \phi)/c - \alpha r \cos(\theta - \phi))d\phi}{[\oint B(\phi)d\phi]^{1/2}[\oint B(\phi)\exp(-2\alpha r \cos(\theta - \phi))d\phi]^{1/2}} \qquad (8)$$

The case of a uniform B, i.e. an isotropic ponderosity, leads to a $C^{(1)}$ that is independent of azimuth $\theta$.  The case is readily evaluated:

$$C^{(1)} = \frac{J_o(\omega r/c + i\alpha r)}{[I_o(2\alpha r)]^{1/2}} \qquad (9)$$

This does not behave like $J_o(\omega r/c) \exp(-\alpha r)$.   For example, it lacks exponential attenuation at large r, and decays geometrically only like $r^{-1/4}$.   Nor does it decay as Tsai [2011] derived for his case of isotropic ponderosity.  His expression $J_o(\omega r/c) / I_o(\alpha r)$ (eqn 25) diminishes asymptotically like $\exp(-\alpha r)$, i.e. exponentially, but not geometrically.   The two expressions differ because of a subtle issue in the notion of uniform *B*.  Uniformity of *B* at the origin is a function of the choice of origin; while Tsai has placed his origin at the midpoint between the stations, here we have placed it at the station at *o*.  The dependence on choice of origin implies that neither expression is very meaningful. This underlines how, as Tsai emphasized, coherency depends on the distribution of sources and is therefore problematic for the retrieval of attenuation.  Inasmuch as the angular distribution of insonifying noise is different at different positions, the azimuthally averaged coherency will also depend on position, even if attenuation α is the same.

For a more arbitrary $B$, $C^{(1)}$ depends on θ. Tsai expands $B$ in a Fourier series and like Prieto calls for an azimuthal average of $C^{(1)}$ over θ. The averaging does not fully simplify the expressions, which remain complex and difficult to interpret. Furthermore, and in spite of the azimuthal averaging, it still depends upon $B(\phi)$. It is therefore sensitive to choice of reference station. Tsai(2011) argues that, for near isotropic ponderosity, the azimuthally averaged coherency approximates his expression (25) which in turn approximates $J_o(\omega r/c)$ exp(-αr). His implication is perhaps that attenuations obtained by comparisons with $J_o$ exp(-αr) for such cases may be valid. This is, however, a very special case, as it requires that the ponderosity be approximately isotropic at the midpoint between the stations. If isotropy obtains instead at one of the stations, the exponential diminishment is lost, as in (9) above. In any case, seismic noise fields are rarely isotropic in practice. Eqn(5), corresponding to a unidirectional ambient noise field and for which $<C^{(1)}>$ has no dependence on α, may more closely correspond to seismic practice. One concludes: coherency $C^{(1)}$ depends on the noise source distribution. Theoretical indications are that even after azimuthal averaging, coherency $C^{(1)}$ cannot reliably be compared to $J_o$ exp(-αr).

The above analysis shows that azimuthally averaged coherency ought not be expected to correspond to a Bessel function times an exponential diminishment exp(-αr) where α is the attenuation. While attenuation can contribute to radial diminishment of the coherency, the effect is not precisely exponential. There are other sources of radial diminishment that can complicate the relation as well. A decayed Bessel function can be expected if wavespeed varies in direction or position. Spatial and/or azimuthal averages may suffer from phase cancellations due to varying wavespeed. This may lead to a diminution of coherency with distance that is unrelated to attenuation α. Consider an oscillatory quantity exp(iωr/c) such as is present in $J_o$, for example coherency in the absence of attenuation α. If this quantity is averaged over a range of wavespeeds c, the result will oscillate and attenuate in emulation of physical attenuation. For purposes of analytic estimation, one may model the variation in slowness 1/c as Gaussian with a mean slowness $\bar{s} = <1/c>$ and a root-variance $\delta s = <1/c^2 - <1/c>^2>^{1/2}$. Then the average of exp(iωr/c) is

$$<\exp(i\omega r/c)> \propto \int \exp(i\omega rs)\exp\{-(s-\bar{s})^2/2\delta s^2\}ds \propto \exp(i\omega r\bar{s})\exp(-\omega^2 r^2 \delta s^2/2) \quad (10)$$

Thus there is a Gaussian diminution with distance r. The diminution is not exponential, but the effect may be, if other uncertainties are present, confused with exp(-αr). It is possible that this has contributed to the apparent attenuation reported by Prieto and co-workers. Lawrence and Prieto (2011), for example, report higher attenuations in the places where Lin et al (2011) and Lin, Ritzwoller and Snieder (2009) report higher wavespeed anisotropies.

Correlation normalized by the mean square field at the reference station

$C^{(2)}$, as defined above is closely related to other work on Green function retrieval from noise correlations. It is, in fact, merely the frequency-domain correlation normalized by the mean square field at the reference station. Matzel [2007] suggested analysis of cross correlations after such normalization for the purposes of retrieving attenuation. As in the previous section, we first ask how C behaves for the special case of a unidirectional noise field, eqn (3). It may be emphasized that, unlike coherency $C^{(1)}$, $C^{(2)}$ depends on site amplification factors; and correspondingly it may not be useful in practice. Nevertheless, leaving aside questions of site amplification factors, the result is

$$C^{(2)} = \exp(i\omega r \cos\theta / c - \alpha r \cos\theta) \qquad (11)$$

On performing an azimuthal average, this becomes

$$<C^{(2)}>_\theta = J_o(\omega r / c + i\alpha r) \qquad (12)$$

The azimuthally averaged $C^{(2)}$ is not $J_o$ times a decaying exponential $\exp(-\alpha r)$. This is readily seen by an analysis of $J_o(z)$'s asymptotic form, which grows with $|z|$ for complex $z$. It is also seen by examination of Fig 1.

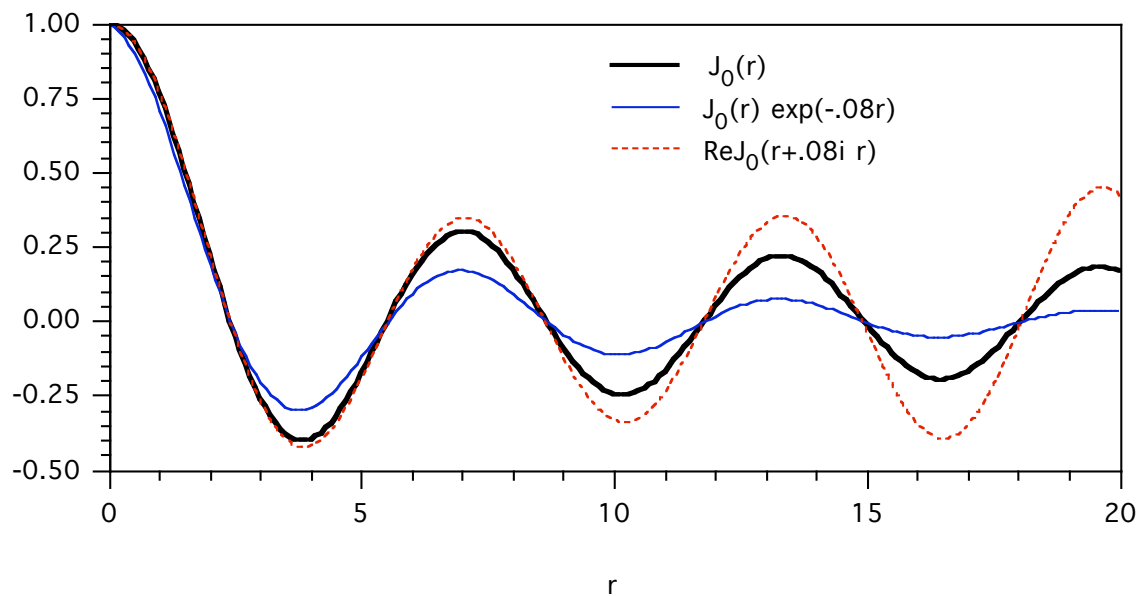

Figure 1] A sample plot of $<C^{(2)}>$ (eqn 12) (red dashed line) for a case in which $\omega/c = 1$ and $\alpha = 0.08$ shows that it grows with distance, faster than a simple Bessel function (bold black line) and much faster than an attenuated Bessel function (solid blue line.)

The result is perhaps surprising. Intuition would maybe have suggested a different behavior. We find here, though, that intuition may be ratified if we further modify the definition, and take only its causal part, the part corresponding to propagation from *o towards x*. Only the stations *x* with positive $r \cos\theta$ have a causal correlation with the station at the origin. We thus restrict the integral over θ to positive cos θ. Then, where the prime ' indicates a zeroing of the anti-causal part,

$$<C^{(2)'}>_\theta = \frac{1}{2\pi} \int_{-\pi/2}^{\pi/2} d\theta \; \exp(i\omega r \cos\theta/c - \alpha r \cos\theta) \quad (13)$$

which may be evaluated asymptotically, for $\omega r/c \gg \exp(2\alpha r)$,

$$<C^{(2)'}>_\theta \sim \frac{1}{\sqrt{2\pi r(i\omega/c - \alpha)}} \exp(i\omega r/c - \alpha r) \quad (14)$$

This is an asymptotic equality, not just a proportionality. The modified average $C^{(2)}$ diminishes with distance as intuited, like $\exp(-\alpha r)/\sqrt{r}$. The asymptotic condition may be relaxed if the end points of the integration are less sharp, i.e, if the restriction to causal correlation (positive lapse time) is imposed smoothly.

If, in addition to the azimuthal averaging, one invokes a spatial average (at fixed separation *r*), there are no differences. Shifts in space are equivalent to changes in A, but the above quantities are independent of A.

It is concluded, at least for this special case in which the ambient field is a plane wave, that i) azimuthally averaged $C^{(2)}$ oscillates like a Bessel function but is *enhanced* by attenuation and distance, and ii) the azimuthally averaged *causal part* of $C^{(2)}$ behaves, at least asymptotically, in the intuited manner: it is diminished by attenuation and distance. These conclusions are independent of choice of origin or spatial averaging.

For a more arbitrary angular distribution of noise intensity $C^{(2)}$ becomes

$$C^{(2)} = \frac{\oint B(\phi) \exp(i\omega r \cos(\theta - \phi)/c - \alpha r \cos(\theta - \phi)) d\phi}{\oint B(\phi) d\phi} \quad (15)$$

This depends on noise directionality B. An average over all azimuth θ at fixed separation *r* and fixed reference station *o*, however, leads to

$$<C^{(2)}>_\theta = J_o(\omega r/c + i\alpha r) \quad (16)$$

for which the factors B drop out. It is concluded: regardless of ponderosity B (and indeed, even if B is highly directional as it would be if the noise field were contaminated by strong point sources), azimuthally averaged $C^{(2)}$ behaves as above, eqn(12); it oscillates like a Bessel function, and attenuation manifests as an *increase* with distance r. It does not decay like $\exp(-\alpha r)$.

A restriction to the causal part of $C^{(2)}$; ie to positive $\cos(\theta-\phi)$, modifies the θ integration.

$$<C^{(2)'}>_\theta = \frac{\frac{1}{2\pi}\int_{\phi-\pi/2}^{\phi+\pi/2} d\theta \oint B(\phi)\exp(i\omega r\cos(\theta-\phi)/c - \alpha r\cos(\theta-\phi))d\phi}{\oint B(\phi)d\phi} \quad (17)$$

The θ integration may be evaluated asymptotically by stationary phase at θ=φ.

$$<C^{(2)'}>_\theta \sim \frac{1}{\sqrt{2\pi r(i\omega/c - \alpha)}} \exp(i\omega r/c - \alpha r) \quad (18)$$

The result finally behaves as originally suggested. This modified quantity oscillates, and decays like exp(-αr)/√r. Again the factors B, and the integration over φ, drop out. Azimuthal averaging has removed the effects of noise directionality.

It is concluded that, while the original coherency prescription $C^{(1)}$ is not reliable, it may be modified. Under circumstances that allow us to neglect site amplification factors or treat them all as identical, the quantity $C^{(2)}$, if azimuthally averaged while restricting to the causal part (and withstanding the complications of phase cancellations due to anisotropic or spatially varying wavespeed) can be fit to a exponentially, exp(-αr), and geometrically, 1/√r, damped oscillation, and attenuation retrieved.

Inasmuch as a shift of origin *o* is equivalent to a change in ponderosity B, and inasmuch as these results are independent of B, they are also independent of spatial averaging.

Spatial and azimuthal averaging in the presence of variable wavespeed, will, as in eqn(10), complicate the analysis. Phase cancellations will cause diminishment with distance *r* unrelated to attenuation. Thus, if one may assume constant site amplification factor, and smooth B, and asymptotic station separations $r \gg c/\omega$, ponderosity B need not be modeled; attenuation can be recovered from azimuthally averaged energy-normalized ray arrival amplitudes. In this author's opinion, these constraints are severe enough that other approaches are worth pursuing[ Weaver 2011b, Lin et al 2012].

Summary

Identification of azimuthally averaged coherency with the intuited $J_o(\omega r/c) \exp(-\alpha r)$ for the retrieval of attenuation α is not justified. Coherency depends too strongly on the directionality of ambient noise intensity – even after coherency is azimuthally averaged. Nor is the convenient assumption that noise is isotropic tenable, as that is an accident of station location. Furthermore, the average coherency $<C^{(1)}>$ can diminish due to phase cancellations related to variations in wavespeed and unrelated to attenuation. It is found, however, that an alternate normalization of correlations can be azimuthally averaged to remove the effects of noise directionality. In the usual asymptotic station separation limit, neglecting differences in site amplification factors, and on restricting it to its causal part, it behaves as intuited.

Aknowledgement   This work was supported by funds from the US DOE through a contract with LANL